# ON THE KINETIC STAGE OF INSTABILITY OF THE ELECTRON BEAM - SOLAR WIND PLASMA SYSTEM

**V.V. Lyahov[1], V.M. Neshchadim**

Institute of Ionosphere, Kamenskoe plato, 050020, Almaty, Kazakhstan

## 1. Introduction

The basic ideas of mechanisms of generation of radio bursts in the solar active regions have developed long ago, actually from the beginning of radio observations of the Sun. It is considered that an electron beam accelerated in the outburst zone excited by rising upwards the oscillations of the environment plasma frequencies of which $\omega_L = \sqrt{\dfrac{ne^2}{\varepsilon_0 m}}$ reduce in the process of decrease of coronal plasma density. Nonlinear interaction causes merging of two longitudinal plasmons into one traversal wave with doubled frequency [1-2]. These transversal waves are radio bursts of III type.

When frequency decreases to several MHz, the ionosphere obtruncates the signal and it becomes inaccessible to recording on the Earth. In the recent decade radio measuring devices were carried out outside the ionosphere, which made it possible to observe radio-frequency radiation of electron beams injected by solar active regions at the distance of one and even two astronomical units [3]. Wavelength of the radio-frequency radiation generated 1a.e. is $\approx$ n·1 km while this value for corona equals to $\approx$ n·10 cm.

This brings up the question about the reason of beam stability as it is known that an electron beam spreading in the background plasma of high density at the over-thermal velocity is unstable [4], and in conditions of solar corona it should quickly dissipate during $\approx$ 0.3 sec., which would not allow it to enter the interplanetary space [5]. Monograph [6] treats possible mechanisms of beam stabilization, including inter conversion of Langmuir longitudinal and traversal electromagnetic waves and nonlinear scattering of a plasma waves excited by a beam on fluctuations of density of the interplanetary plasma leading to their coming out of resonance with beam because of change of phase velocity of the wave.

The theory based on the idea of interchange in space of generation zone and "silent" zones (in these zones the wave comes out of resonance with beam) was called a "theory of stochastic generation" [7,8] and was applied to the problem of radio-frequency radiation on a near-Earth shock wave [9]. However, a consistent quantitative theory of radio bursts of III type in an interplanetary space satisfying the whole complex of measurements still not developed. Various problems of nonlinear stage of radio-frequency generation are being analyzed: the cascade theory [10], a problem of restriction of amplitude wave growth with distribution function going on "plateau" [7], etc., but dispersion properties of oscillating beam-plasma system still insufficiently studied.

---

[1] Corresponding author, mail: v_lyahov@rambler.ru

Our previous papers first of all analyzed frequency characteristics of perturbations of the system of electron beam – solar wind plasma in the linear approximation [11,12] and at the second investigated hydrodynamic stage of beam instabilities in the quasilinear approximation [13].. This paper investigates the kinetic instability of the beam - plasma system that developes after the hydrodynamic stage.

## 2. Statement of the Problem

We believe that energy resulting from the development of instability of oscillations is small in comparison with thermal energy of plasma particles, i.e. plasma is weakly turbulent:

$$\frac{W}{n\varepsilon} << 1, \tag{1}$$

Here, $n\varepsilon$ - thermal energy of particles, $W = \sum_k W_k = \sum_k \frac{\varepsilon_0 E_k^{\,2}}{2}$ - energy of plasma oscillations of unit volume, $n$ - concentration of particles.

Energy of plasma oscillations exceeds energy of thermal fluctuations of an electromagnetic field in plasma [14]:

$$\frac{W}{n\varepsilon} >> (\frac{e^2 n^{\frac{1}{3}}}{4\pi\varepsilon_0 \varepsilon})^{\frac{3}{2}}. \tag{2}$$

Within the limits of approximation (1), (2) hydrodynamic instability of the system of electron beam – solar wind plasma has been explored and it became clear that at the hydrodynamic stage of beam energy loss is insignificant and mainly goes into wave excitation and to a lesser degree into beam heating, and the time of development of hydrodynamic instability is [13]:

$$t_{_{Max}} = \frac{mu^2}{2D}(\frac{n_b}{2n_e})^{\frac{2}{3}} \approx 10^4 c. \tag{3}$$

Here, $m$ – electron mass; $u$ – beam velocity; $n_b$ and $n_e$ – density of electrons in a beam and solar wind plasma, respectively; $D$ – electron diffusion coefficient in velocity space.

Using results of the study of hydrodynamic instability, we will analyze the main parameters of further developing kinetic stage of beam instability.

## 3. Study of the kinetic stage of instability of the system of electron beam – solar wind plasma in the quasilinear approximation.

Hydrodynamic instability leads to thermalization of beam electrons, i.e. to dithering of their distribution function. At the end of development of hydrodynamic instability ($t_{_{Max}} \approx 10^4 c$), the kinetic instability leading in the long run to formation of a horizontal plateau in the beam zone in the tail-end of Maxwellian allocation begins to develop.

Let's trace behavior of the beam at the kinetic stage of instability and consider relaxation of diffusion electron beam in solar wind plasma. It may be thought that at the initial moment of development of the kinetic stage of instability a full function of electron distribution in the beam-plasma system looks like:

$$F_0(v,0) = \left(\frac{m}{2\pi}\right)^{\frac{1}{2}} \frac{1}{\sqrt{T_0(r_0)}} \exp(-\frac{mv^2}{2T_0(r_0)}) + \frac{N_b(r_0)}{N_0(r_0)}\left(\frac{m}{2\pi T_b(r_0)}\right)^{\frac{1}{2}} \exp(-\frac{m(v-u)^2}{2T_b(r_0)}), \tag{4}$$

where

$N_0(r_0), N_b(r_0), T_0(r_0), T_b(r_0)$ - values of density of electrons of background solar wind plasma and beam, and also corresponding temperatures at $r = r_0$ point of the interplanetary space which has been reached by the electron beam moving at $u$ velocity during time $t_{max} \approx 10^4 c$. Fig. 1 shows diagram of distribution function (4). On the abscissa velocity $v$ is plotted and on ordinate a distribution function $F_0(v,0)$ is plotted.

The system of quasilinear equations for kinetic instability can be derived much as it has been made in the study of hydrodynamic stage [13]:

$$\frac{\partial F_0}{\partial t} = \frac{\partial}{\partial p} D \frac{\partial F_0}{\partial p},$$

$$\frac{\partial |E_k|^2}{\partial t} = 2\delta_k |E_k|^2,$$  (5)

$$D = \frac{e^2}{2} \frac{|E_k|^2}{kv}.$$

Here, $E_k$ – electric field strength of plasma oscillations at the kinetic stage, $\delta_k$ - increment of kinetic instability development, $k$ – wave number of the appropriate plasma mode.

It has been shown [14] that the system (5) allows for the existence of stationary solution in which $\left.\frac{\partial F_0(v,t)}{\partial v}\right|_{v=\frac{\omega}{k}} \to 0$, i.e. the distribution function in some interval of velocities in the vicinity of point $v = \frac{\omega}{k}$ forms a plateau.

The set of equations (5) makes it possible to determine evolution of the distribution function $F_0(v,t)$ from the initial configuration to the final configuration (Fig. 1: "trough" and "hump" are replaced with a horizontal line). The problem of quasilinear theory lies in the definition of values of velocity $v_1$ and $v_2$ at which this plateau begins and ends, i.e. abscissas of intersection points of the initial curve and formed plateau as well as ordinate of this plateau $F_0(v,\infty)$.

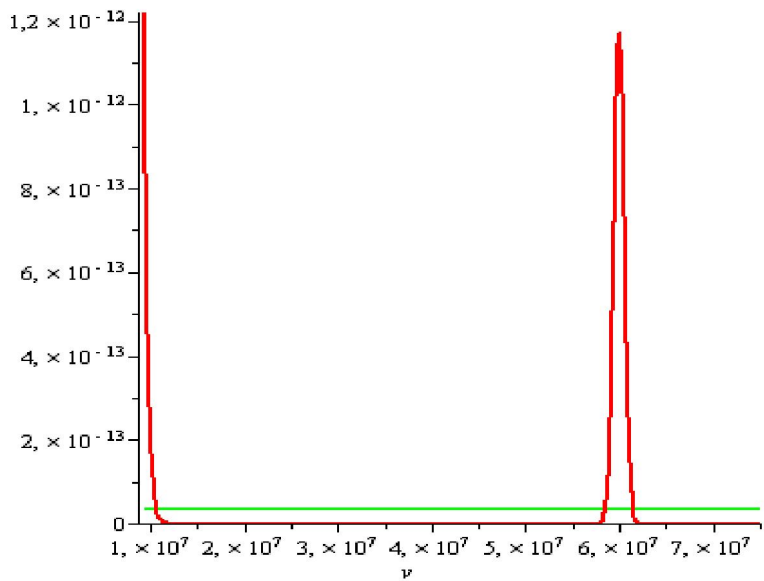

Fig. 1

Three values $v_1$, $v_2$ and $F_0(v, \infty)$ can be derived from relations:

$$\int_{v_1}^{v_2} F_0(v,0)dv = F_0(v,\infty)(v_2 - v_1),$$  (6)

$$F_0(v_1,0) = F_0(v,\infty),$$  (7)

$$F_0(v_2,0) = F_0(v,\infty).$$  (8)

Equation (6) represents the particle number conservation law. Relations (7), (8) are obvious. Set of equations (6), (7), (8) was solved numerically in MAPLE 13 package. Thus:

$$v_1 = 1,07 \cdot 10^7 \text{ Km/s, km/s } v_2 = 6,15 \cdot 10^7, \text{ and } F_0(v,\infty) = 3,29 \cdot 10^{-14} \text{ м}^{-3} \text{ с}^{-1}.$$  (9)

In solving equations (6), (7), (8) values of the following constants have been used:

$$m = 9,1 \cdot 10^{-31} кг,$$
$$u = 0,6 \cdot 10^8 м/с,$$
$$r_0 = u \cdot t_{max} = 6,0 \cdot 10^{11} м = 4 a.е.,$$
$$T_0(r_0) = 3,3 \cdot 10^{-18} Дж,$$  (10)
$$T_b(r_0) = 2,9 \cdot 10^{-19} Дж,$$
$$N_0(r_0) = 0,3 \cdot 10^6 м^{-3},$$
$$N_b(r_0) = 5,0 \cdot 10^{-1} м^{-3}.$$

As it has been already determined [14], in our case:

$$\frac{|E_k(\infty)|^2}{N_b \frac{mu^2}{2}} \approx \frac{v_2 - v_1}{u} = 0.84.$$  (11)

That is, when quasilinear relaxation of the system of electronic beam – solar wind plasma at kinetic stage of instability develops, more than a half of kinetic energy of a beam is transmitted to plasma oscillations.

## 4. Conclusion

Within the framework of quasilinear theory it has been found that when radio bursts of III type are generated by the inhomogeneous system of electron beam – solar wind plasma, only time of development of the primary stage of instability, i.e., agnetohydrodynamic, is more than the total time of relaxation of an electron beam derived within the homogeneous model. During the development of hydrodynamic instability the electron beam is spread from the Sun to the distance of 4 astonomical unities. the hydrodynamic stage is followed by the development of kinetic instability leading, ultimately, to the formation of horizontal plateau in the beam region at the tail of Maxwell distribution. Parameters of the plateau, such as its length and height, have been calculated as characteristic parameters of an electron beam generated by the active solar bursts. In the development of kinetic beam instability more than a half of kinetic energy of the beam is transmitted to plasma oscillations.

## REFERENCES

1.   Parker E.N. Astrophys. J. 1959. Vol.129. p.217.

2.   Ginzburg V.L., Zheleznyakov V.V. Astron. J. 1958. Vol. 35. p. 691. (in Russian)

3.   Benz A.O., Grigis P.C., Csillaghy A. 2005. Solar Phys. Vol. 226. pp. 121-142.

4.   Vedenov A.A., Velikhov E.P., Sagdeyev R. Z. Nuclear Synthesis, 1, 82, 2, 465. (in Russian)

5.   Kaplan S.A., Tsytovich V. N. Plasma Astrophysics.1972. Science. M. 270p. (in Russian).

6.   Kaplan S.A., Pikelner S.B., Tsytovich V. N. Plasma physics of Solar atmosphere. 1977. Science. M. 256 p.  (in Russian)

7.   Robinson R.A., Cairns I.H., Gurnett D.A. 1992. Astrophys. J. Vol. 387. pp. L101-L104.

8.   Robinson R.A., Cairns I.H. 1994. Solar. Phys. v.154. p.335-360.

9.   Kuncic Z., Cairns I.H., Knock S.A. 2004. J. Geophys. Res. Vol. 109. CiteID A02108.

10.  Kontorovich V. M, Pimenov S.V., Tsvyk N.A. Astron. J. 1993. Vol. 70. iss. 3. pp. 571-582. (in Russian)

11.  Lyahov V.V. Space Explorations. 2008. Vol. 46. No. 5, pp. 412-417. (in Russian)

12.  Lyahov V.V. Space Explorations. 2009. Vol. 47, N 1. pp. 33-37. (in Russian)

13.  Lyahov V.V., Neshchadim V.M. Planetary and Space Science.2009.Vol. 57. pp. 415-423.

14.  Aleksandrov A.S., Bogdankevich L.S., Ruhadze A.A. Essentials of Plasma Electrodynamics. 1978. Higher School. M. 407p. (in Russian)